\documentclass[10pt,letter]{article}

\usepackage{hyperref}
\usepackage[numbers]{natbib}
\usepackage{times}

\usepackage{mathrsfs}
\usepackage{graphicx}
\usepackage{amsmath} 
\usepackage{amsfonts} 
\usepackage{amssymb}
\usepackage[utf8]{inputenc}

\newcommand{\itm}{\textsl{ITM Probe}}
\newcommand{\ssum}{\textsl{SaddleSum}}
\newcommand{\cssm}{\textsl{CytoSaddleSum}}
\newcommand{\citm}{\textsl{CytoITMprobe}}
\newcommand{\qmbpmnt}{\textsl{qmbpmn-tools}}

\newcommand{\mat}[1]{\mathbf{#1}}
\newcommand{\PhiT}[3]{\varPhi_{#1,#3}^{#2}}
\newcommand{\nPhiT}[3]{\hat{\varPhi}_{#1,#3}^{#2}}
\newcommand{\I}{\mathbb{I}}

\begin{document}

\begin{titlepage}

\begin{center}
{\Large\bf CytoITMprobe: a network information flow plugin for Cytoscape}
\end{center}
\vspace{.35cm}

\begin{center}
{\large Aleksandar Stojmirovi\'c\,, Alexander Bliskovsky and Yi-Kuo Yu\footnote{to whom correspondence should be addressed}}
\vspace{0.25cm}
\small

\par \vskip .2in \noindent
National Center for Biotechnology Information\\
National Library of Medicine\\
National Institutes of Health\\
Bethesda, MD 20894\\
United States
\end{center}

\normalsize
\vspace{0.25cm}

\begin{abstract}

\subsubsection*{Background:} Cytoscape is a well-developed flexible platform for visualization, integration and analysis of network data. Apart from the sophisticated graph layout and visualization routines, it hosts numerous user-developed plugins that significantly extend its core functionality. Earlier, we developed a network information flow framework and implemented it as a web application, called \itm. Given a context consisting of one or more user-selected nodes, \itm\ retrieves other network nodes most related to that context. It requires neither user restriction to subnetwork of interest nor additional and possibly noisy information. However, plugins for Cytoscape with these features do not yet exist. To provide the Cytoscape users the possibility of integrating \itm\ into their workflows, we developed \citm, a new Cytoscape plugin. 

\subsubsection*{Findings:} \citm\ maintains all the desirable features of \itm\ and adds additional flexibility not achievable through its web service version. It provides access to \itm\ either through a web server or locally. The input, consisting of a Cytoscape network, together with the desired origins and/or destinations of information and a dissipation coefficient, is specified through a query form. The results are shown as a subnetwork of significant nodes and several summary tables. Users can control the composition and appearance of the subnetwork and interchange their \itm\ results with other software tools through tab-delimited files.

\subsubsection*{Conclusions:} The main strength of \citm\ is its flexibility. It allows the user to specify as input any Cytoscape network, rather than being restricted to the pre-compiled protein-protein interaction networks available through the \itm\ web service. Users may supply their own edge weights and directionalities. Consequently, as opposed to \itm\ web service, \citm\ can be applied to many other domains of network-based research beyond protein-networks. It also enables seamless integration of \itm\ results with other Cytoscape plugins having complementary functionality for data analysis.
\end{abstract}
\end{titlepage}

\section*{Background}

Cytoscape~\cite{CSCK07,SMOB03,SORW11} is a popular and flexible platform for visualization, integration and analysis of network data. Apart from the sophisticated graph layout and visualization routines, its main strength is in providing an API that allows developers other than its core authors to produce  extension plugins. Over the last decade, a large number of plugins have been released, supporting the features such as import and export of data, network analysis, scripting and functional enrichment analysis. In this paper, we describe \citm\, a plugin that brings to Cytoscape new functionality founded on information flow.

Numerous approaches for analyzing biological networks based on information flow~\cite{NJAC05,TWAC06,SBKE08,ZMOP08,MLZR09,VTX09,KPWP11} have emerged in recent years. The main assumption of all such methods is information transitivity: information can flow through or can be exchanged via paths of biological interactions. Our contribution to this area~\cite{SY07,SY10b} is a context-specific framework based on discrete-time random walks (or equivalently, diffusion) over weighted directed graphs. In contrast to most other approaches, our framework explicitly accommodates directed networks as well as the information loss and leakage that generally occurs in all networks. Apart from the network itself and a user-specified context, it requires no prior restriction to the sub-network of interest nor additional and possibly noisy information. We implemented our framework as an application called \itm~\cite{SY09b} and made it available as a web service~\cite{ITMsite}, where users can query protein-protein interaction (PPI) networks from several model organisms and visualize the results.

In addition to implementing network flow algorithms, the \itm\ web service possesses a number of useful features. Using the Graphviz~\cite{GN00} suite for layout and visualization of graphs, it displays in the user's web browser the images of subnetworks consisting of nodes identified as significant by the information flow models and offers a choice of multiple coloring schemes. The entire query results can be retrieved in the CSV format or forwarded to a functional enrichment analysis tool to facilitate their interpretation. However, lacking a mechanism to decouple the algorithmic part from the interaction graph, the \itm\ web service restricts users to querying only the few compiled PPI networks available on the website. Using a canned suite for graph layout, \itm\ limits the users' ability to manipulate network images. For example, the only way to change the layout of significant subnetworks is to choose a different seed and re-compute the layout. Most importantly, not having an adequate interface to a well-designed platform such as Cytoscape, it is difficult to use the results of the \itm\ service within the workflows involving additional data and algorithms from other sources. We thus developed \citm\ to meet these challenges by (1) providing an explicit decoupling between the algorithmic part and the interaction graph, (2) utilizing the core graph manipulation functionality of Cytoscape for a broader visualization choices, and (3) adding an appropriate input/output interface for seamless integration with other resources available in Cytoscape.

\begin{figure*}[h!]
\begin{center}
\scalebox{0.8}{\includegraphics{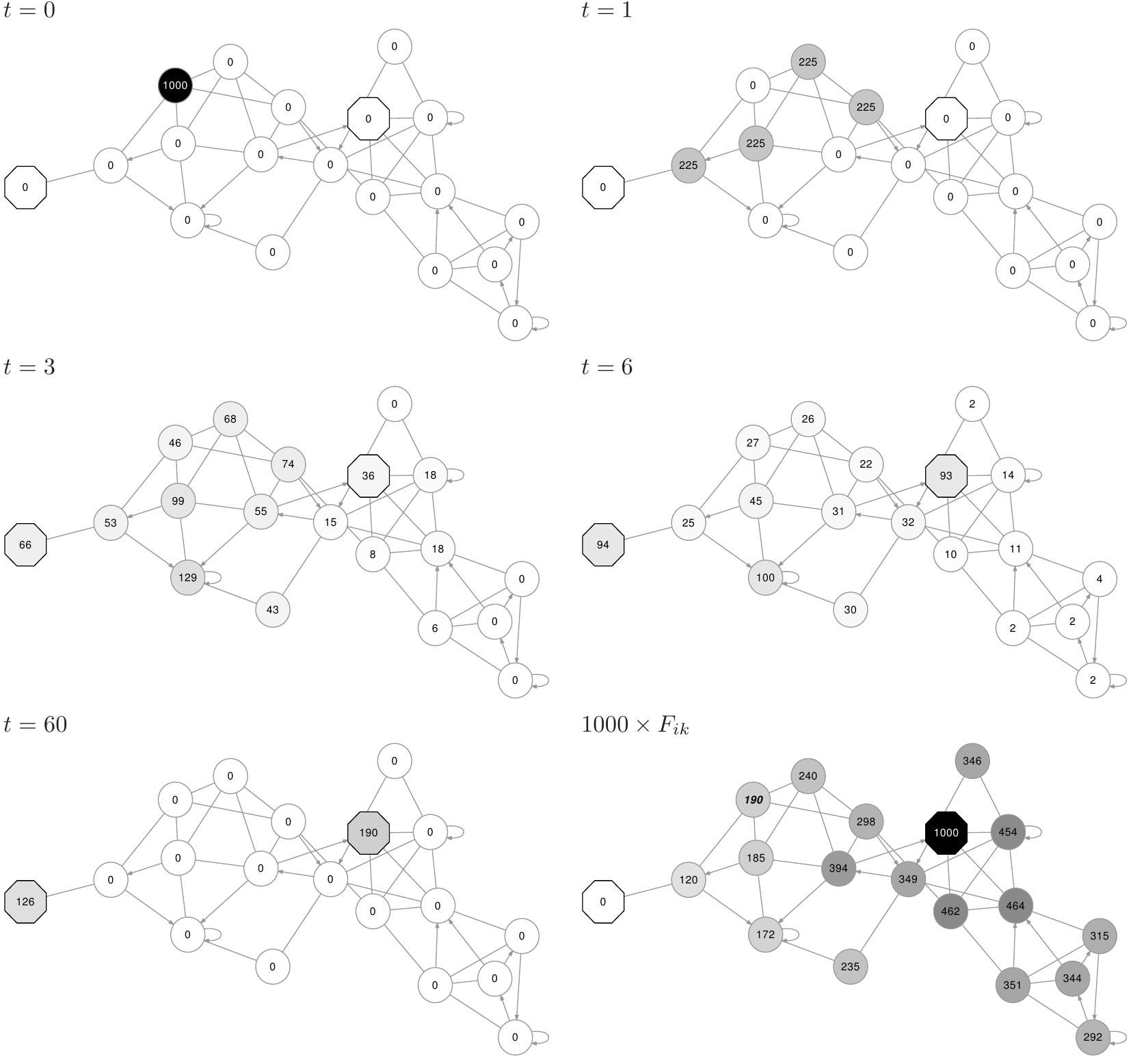}}
\caption{\textbf{ITM Probe is based on discrete-time random walks with boundary nodes and damping.} As an example, consider the weighted directed network shown, containing 19 nodes and 44 links. Single-directional links are assigned weight 2 and are indicated using arrows while bi-directional edges are assigned weight 1 and are shown as lines. The first five graphs show the time progress of a random walk in the presence of damping and two absorbing boundary nodes (indicated by octagons). At $t=0$, 1000 random walkers start at a single point in the network. At $t=1$, they have progressed one step from their origin to the nodes adjacent to it, being distributed randomly in proportion to the weights of the edges leading from the origin. Only 900 walkers remain in the network at $t=1$ due to damping: the damping factor $\mu=0.9$ (dissipation $0.1$) means that $10\%$ of walkers are dissipated at each step. At $t=60$, most of the walks have terminated, either by dissipation, or by reaching one of the two boundary nodes. The number of walkers terminating at each boundary node depends on their starting location. The final graph shows the probability $F_{ik}$ for a random walk starting at any transient node in the network (indicated by circular shape) to terminate at the boundary node on the right-hand side (scaled by 1000). Note that the value indicated in the final graph for the starting node at $t=0$ (190) is the same as the final number of walks shown at $t=60$ as terminating at the right boundary node.} \label{fig:absorbing}
\end{center}
\end{figure*}

\section*{Information Flow Framework}

\itm\ extracts \emph{context-specific} information from networks. We elaborated on the information flow framework underlying \itm\ in our previous publications~\cite{SY07,SY10b} and here we provide a non-technical explanation.  Given a context consisting of one or more user-selected network nodes, the aim is to retrieve a set of other network nodes most related to that context. We model networks as weighted directed graphs, where nodes are linked by directional edges and each edge is assigned a positive weight. One can consider a random walker that wanders among network nodes in discrete steps. The rule of the walk is that the walker starts at a certain node and in each step moves randomly to some adjacent node with probability proportional to the weight of the edge linking these nodes (Fig.~\ref{fig:absorbing}). If the graph is connected, that is, if there is a directed path linking any two nodes, such a walk never terminates and the walker will eventually visit every node in the graph.

Our main idea is to set termination or \emph{boundary} nodes for the walkers while using random walks to explore the neighborhoods of the context nodes. Provided there is a directed path linking any node to a boundary node, every random walk here will eventually terminate. Furthermore, the nodes visited by a walker before termination will vary depending on the origin of the walk. Since a random walk is a stochastic process, and each walk is different, we are interested in the cumulative behavior of infinitely many walkers following the same rules. On average, we expect that the nodes more relevant to the context will be more visited than those that are less relevant. Thus, the main quantity of interest is the average number of visits to a node given the selected origins and destinations of the walk.

A problem with the above approach is that random walkers may spend too much time in the graph if the origins and destinations of the walk are far apart. This could mean that the entire graph is visited so that the most significant nodes are just those with the largest degree. To ensure that the significant nodes are relatively close to the context nodes, our framework contains an additional ingredient, \emph{damping}: at each step of a walk, we assign a certain probability for the walker to dissipate, that is, to leave the network. We still evaluate the average number of visits to each node, but now only count the visits prior to the walker leaving the network. Evidently, the nodes that are close to the walker's origin will be significantly visited. In addition to forcing locality, damping is also natural in physical or biological contexts. If we treat random walkers as information propagating through the network, it is natural to assume that some information is lost during transmission. For protein-protein interaction networks, where nodes are proteins and links are physical bindings between proteins, damping could be associated with protein degradation by proteases, which would diminish the strength of information propagation.

\begin{figure*}[h!]
\begin{center}
\scalebox{0.8}{\includegraphics{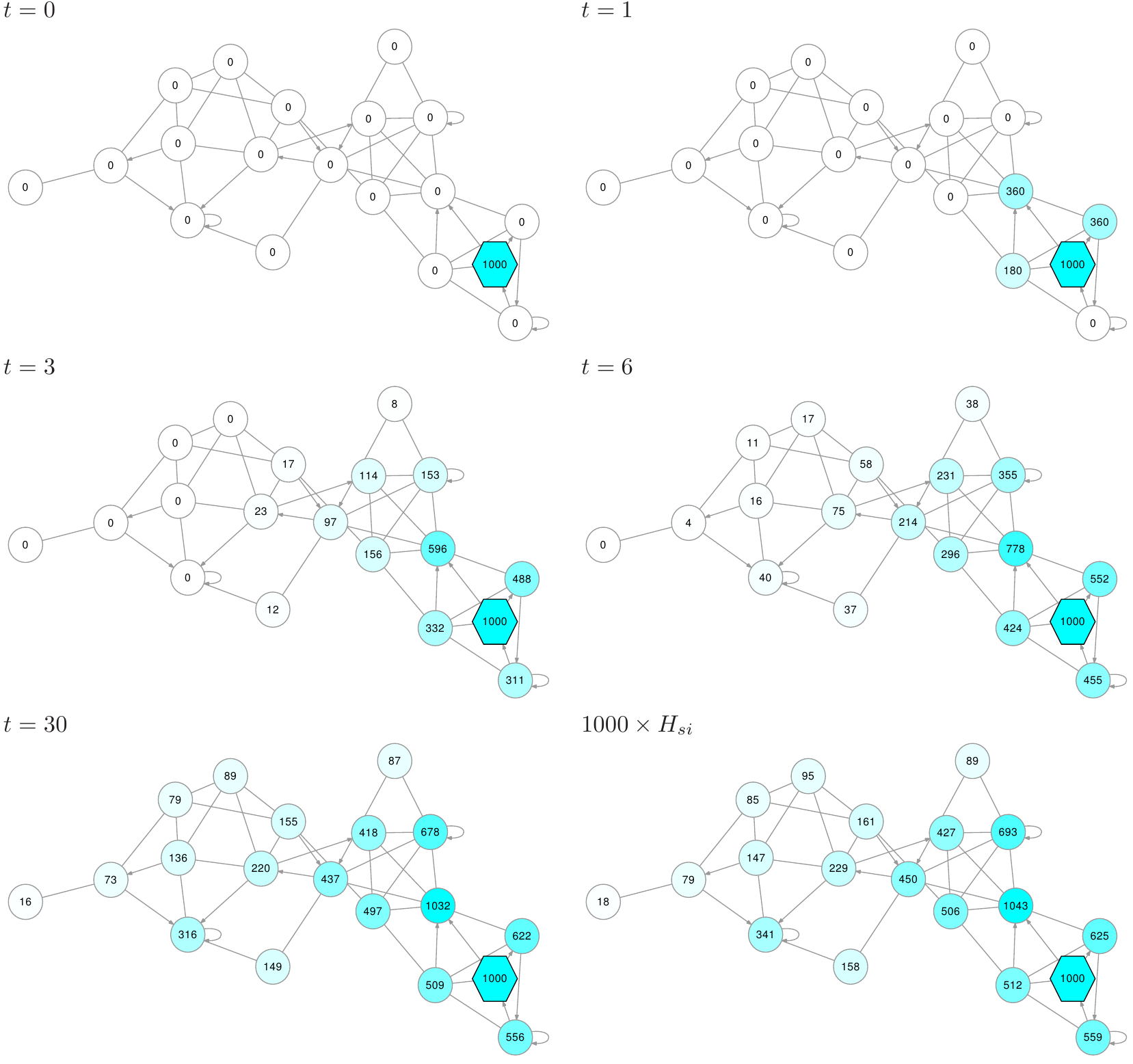}}
\caption{\textbf{The emitting model counts visits from sources.} Using the example network from Fig.~\ref{fig:absorbing} with the same damping factor, consider the case where 1000 random walkers start at the source node indicated by a hexagon. At each time step, some random walkers leave the network due to damping or by moving back to the source. In the first five graphs, the number in each node documents the total number of visits to that node from all random walkers, dissipated or not, up to the indicated time. The value of $H_{si}$ returned by the ITM Probe emitting mode ($s$ here denotes the source node) yields the expected number of visits to node $i$ per random walker that starts at $s$ over infinitely many time steps. The final graph shows the values of $H_{si}$ for this context, scaled by 1000. Note that the magnitude shown for one transient node is greater than 1000 because a walker may visit the same node multiple times.} \label{fig:emitting}
\end{center}
\end{figure*}

\itm\ framework contains three models: \emph{emitting}, \emph{absorbing} and \emph{channel}. In the absorbing model (Fig.~\ref{fig:absorbing}), the context nodes are interpreted as destinations or \emph{sinks} of random walks, while every non-boundary or \emph{transient} node is considered as a potential origin. For each transient node $i$ and each sink $k$, the model computes $F_{ik}$, the average number of visits to the terminating node $k$ by random walks originating at the node $i$. Since a walk can either terminate at one sink or the other, $F_{ik}$ can also be interpreted as the probability that a random walk from $i$ reaches $k$. In the absence of damping, the sum of $F_{ik}$ over all sinks will be exactly $1$ for any transient node $i$. However, in the presence of damping, the sum of $F_{ik}$ over all sinks may be much less than $1$ (Fig.~\ref{fig:absorbing}). The emitting model (Fig.~\ref{fig:emitting}), offers a dual point of view. Here, the context nodes are interpreted as origins or \emph{sources} of random walks. The walks terminate by dissipating or by returning to the sources -- the sources form an emitting boundary. Since the origins of the walks are fixed, the quantity of interest is the visits to the transient nodes. Specifically, for each source $s$ and each transient node $i$, the emitting model returns $H_{si}$, the average number of visits to $i$ by walkers originating at $s$. 

The values of $F_{ik}$  and $H_{si}$ can be efficiently computed by solving (sparse) systems of linear equations. Let $W_{ij}$ denote the weight of the directed link $i\to j$ and let $0<\mu<1$ denote the damping factor. For all pairs of nodes $i,j$, construct the random walk evolution operator $\mat{P}$, where $P_{ij} = \frac{\mu W_{ij}}{\sum_{j'} W_{ij}}$. The operator $\mat{P}$ includes damping and hence $\sum_{j} P_{ij} < 1$. Let $\mat{P}_{TT}$ denote the sub-operator of $\mat{P}$ with domain and range restricted only to transient nodes and let $\mat{G} = (\I - \mat{P}_{TT})^{-1}$, where $\I$ stands for the identity matrix. Then, it can be shown~\cite{SY07}, that 
\begin{align*}
F_{ik} &= \sum_{j} G_{ij} P_{jk}, \qquad \text{and} \\
H_{si} & = \sum_{j} P_{sj} G_{ji}.
\end{align*}
More details, including the cases where $\mu=0$, $\mu=1$ or non-uniform damping are covered in \cite{SY07,SY10b}.

\begin{figure*}[h!]
\begin{center}
\scalebox{0.8}{\includegraphics{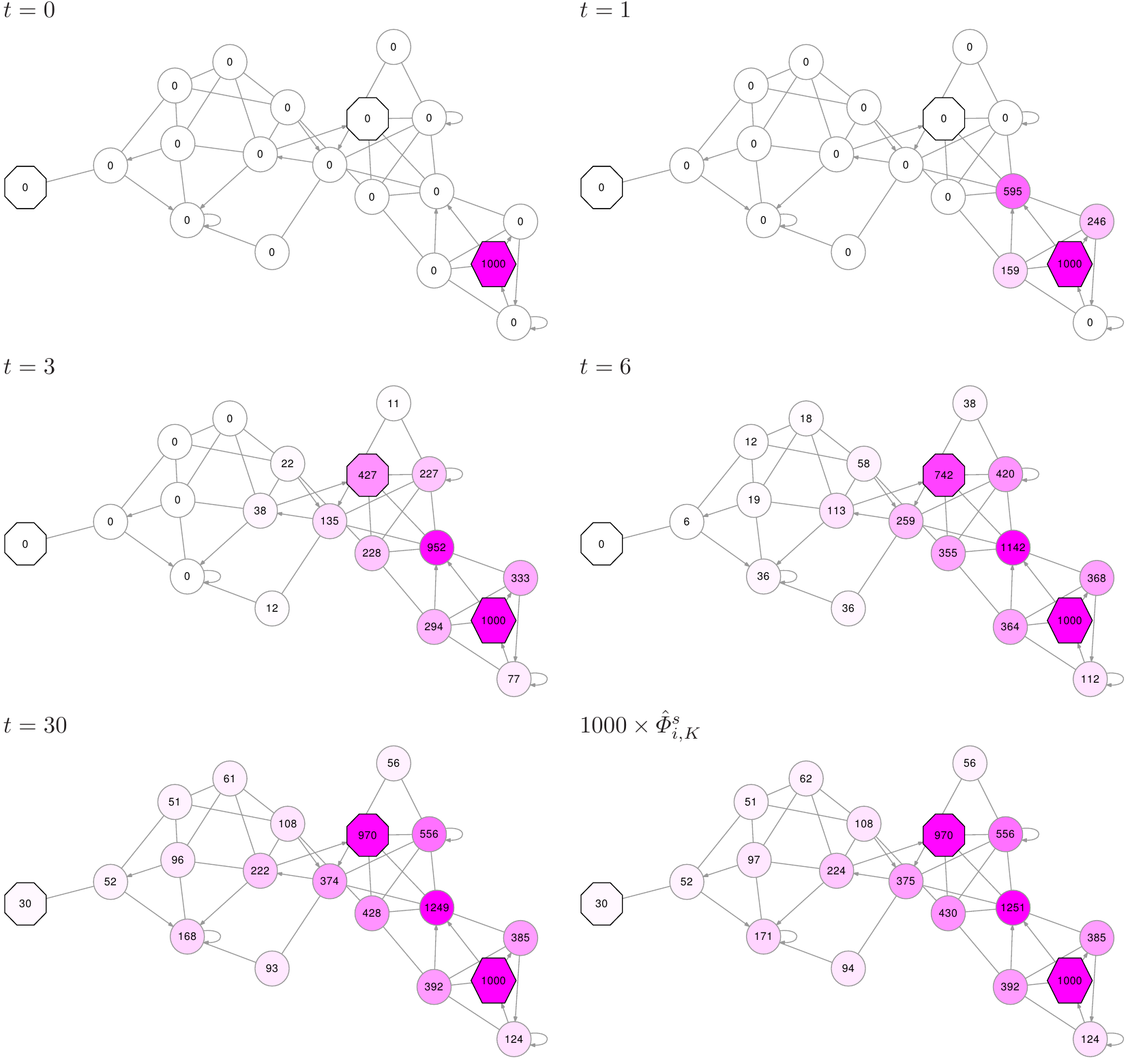}}
\caption{\textbf{The channel model highlights the directed flow from origins to destinations.} Consider once again the example network from Figs.~\ref{fig:absorbing} and \ref{fig:emitting}, now with a single source (hexagon) and two sinks (octagons). In common with the case from Fig.~\ref{fig:emitting}, the walkers start at the source, but in this case can terminate only by reaching the sinks. The damping factor is implicit: it determines how far the walkers are allowed to deviate from the shortest path towards one of the sinks. In the first five graphs, the number in each transient node documents the total number of visits to that node from all random walkers up to the indicated time. However, the value in each sink node represents the likelihood to reach that sink from the source at the indicated time. The value of $\nPhiT{i}{s}{K}$ returned by the ITM Probe normalized channel mode yields the expected number of visits to node $i$ per random walker that starts at $s$ over infinitely many time steps. Note that the sink nodes split the flow from the source depending on their location. In this example, over infinitely many time steps, the node closer to the source captures 970 walkers, while the further sink gets only the remaining 30.} \label{fig:channel}
\end{center}
\end{figure*}

\begin{figure}[h!]
\begin{center}
\scalebox{0.88}{\includegraphics{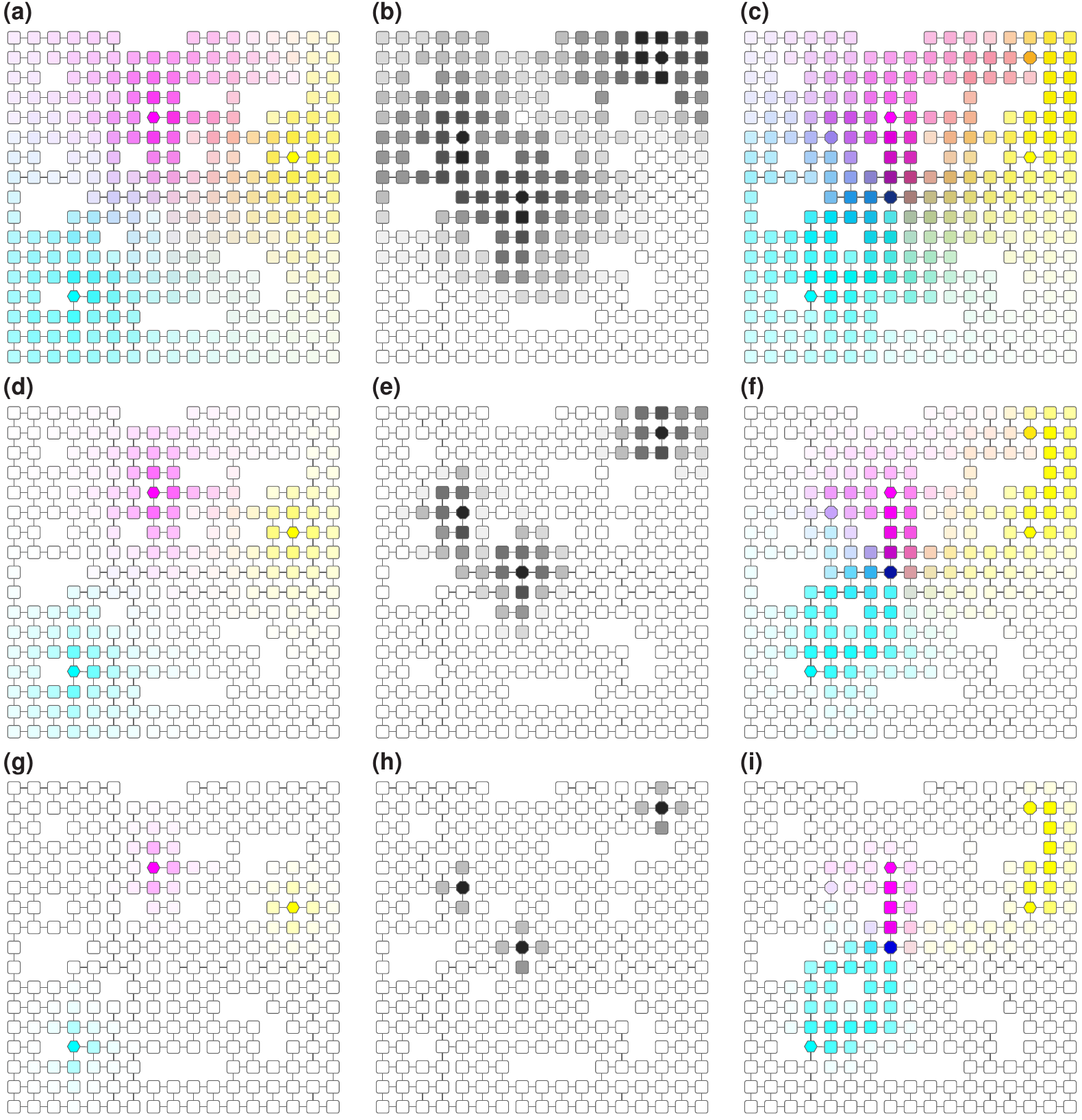}}
\caption{\textbf{An example of the results of running different \itm\ models.}
Here we see the results of running the emitting (a,d,g), absorbing (b,e,h) and channel (c,f,i) model of \itm\ with the same sources and sinks but different dissipation coefficients. The underlying undirected graph is derived from a square lattice by removing random nodes and edges. Sources are shown as hexagons, sinks as octagons, and transient nodes as squares. The top row (a,b,c) shows the runs with damping factor $\mu=0.95$ (dissipation $0.05$), the middle (d,e,f) with $\mu=0.75$ and the bottom with $\mu=0.25$. For the emitting and channel model, each basic cyan, magenta or yellow color is associated with a source. The coloring of each node arises by mixing the basic color in proportion to the strength of information flow from their respective sources. For the absorbing model, the nodes are shaded according to the total probability of absorption at any sink on a logarithmic scale. 
} \label{fig:models}
\end{center}
\end{figure}

The channel model combines the emitting and the absorbing model, with both sources and sinks on the boundary. It illuminates the most likely paths from sources to sinks. For each source node $s$, transient node $i$ and sink node $k$, it computes $\PhiT{i}{s}{k}=H_{si}F_{ik}$, the average number of visits to $i$ by a random walker that originates at $s$ and terminates at $k$. \itm\ does not report $\PhiT{i}{s}{k}$ directly, but instead shows a simpler, \emph{normalized} quantity $\nPhiT{i}{s}{K}$ (Fig.~\ref{fig:channel}), which is defined for each source $s$ and transient node $i$ by
\begin{equation}
\nPhiT{i}{s}{K} = \frac{\sum_{k} H_{si}F_{ik} }{\sum_{k'} F_{sk'}}.
\end{equation}
Here, the numerator $\sum_{k} H_{si}F_{ik}=\sum_k \PhiT{i}{s}{k}$ gives the average number of visits, in the presence of damping, to $i$ by a random walker starting at $s$ and terminating at any sink. The denominator gives the total probability of a walker starting at $s$ to terminate at any sink. Hence, with the denominator off-setting the effect of damping, 
the value of $\nPhiT{i}{s}{K}$ counts the average number of visits to $i$ by walkers that start at $s$ and terminate at any of the sinks as if no dissipation is present. Generally, damping in the emitting or the absorbing model determines how far the flow can reach away from its origins. In contrast, the damping parameter for the normalized channel model plays a different role  (Fig.~\ref{fig:models}): it effectively determines the `width' of the channel from sources to sinks. When damping is very strong, only the nodes on the shortest path from a source to its nearest sink will be visited.

Given the close relationship between random walks and diffusion, it is also possible to interpret \itm\ models through information diffusion (or information flow). Within that paradigm, a fixed amount information is constantly replenished at the source nodes while leaving the network at all boundary nodes and through dissipation. At equilibrium, when the rate of flow entering equals the rate of leaving, the amount of information occupying each transient node is equivalent to the average number of visits to that node (using the aforementioned non-replenishing random walk interpretation~\cite{SY07}). We call the set of nodes most influenced by the flow an \emph{Information Transduction Module} or ITM.

%%%%%%%%%%%%%%%%%%%%%%%%%%%%%%%%
\section*{Software architecture}

\begin{figure}[ht!]
\begin{center}
\includegraphics{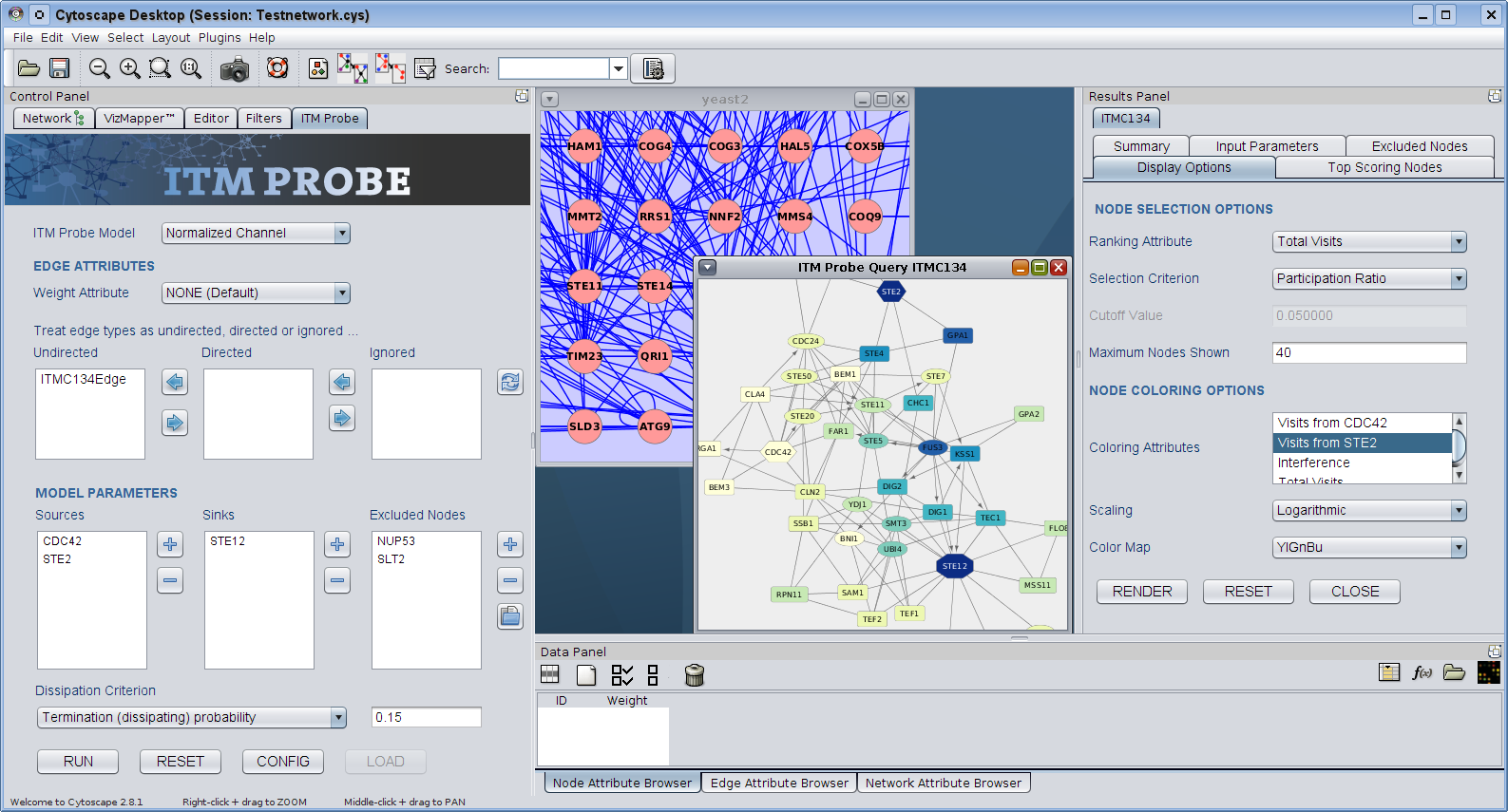}
\caption{\textbf{\citm\ interface.}
At startup from the Plugins menu, \citm\ embeds its query form into the Control Panel (left). After performing a query or loading previously obtained search results, it creates an ITM subnetwork showing significant nodes and a viewer embedded into Results Panel (right). The viewer allows closer examination of the results and manipulation of the contents and the look of the ITM subnetwork. The overall visual styling of \citm\ components closely resembles that of the \itm\ web version.} \label{fig:screenshot}
\end{center}
\end{figure}

\citm\ architecture consists of two parts: the user interface front end and computational back end. The user interface, written in Java~\cite{Java} using Cytoscape API, is accessed as a Cytoscape plugin. It consists of the query form, the results viewer and the ITM subnetwork (Fig.~\ref{fig:screenshot}). The back end is the standalone \itm\ program, written in Python, which can be installed locally or accessed through a web service. In either configuration, \citm\ takes the user input through the graphical user interface, validates it, and passes a query to the back end. Upon receiving from the back end the entire query results, \citm\ stores them as the node and network attributes of the original network. Consequently, the query output can be edited or manipulated within Cytoscape, as well as saved for later use.

Standalone \itm\, is a part of the \qmbpmnt\ Python package, which also contains the code supporting the \itm\ and \ssum\ web services, as well as the scripts for constructing the underlying datasets. The \itm\ part depends on Numpy and Scipy~\cite{JOPo01} packages for numerical computations. The performance of \itm\ critically depends on the routines for computing direct solutions of large, sparse, nonsymmetric systems of linear equations. Scipy supports two sparse direct solver libraries (both written in C): SuperLU~\cite{DEGX99} as default and UMFPACK~\cite{Davis04} as an optional add on through SciKits collection~\cite{Scikits}. In our experience, UMFPACK runs faster than SuperLU and Scipy always uses it if available. However, for optimal performance, UMFPACK requires well-tuned Basic Linear Algebra Subroutines (BLAS) libraries and may not be easy to install. To support users who are inclined not to install UMFPACK or Scipy, \citm\ supports remote queries by default. 

\section*{Input}

\citm\ requires as input a weighted directed graph and the \itm\ model parameters that include a selection of boundary nodes and a dissipation probability. 

\subsection*{Step one: defining a query graph}

In \citm\, graph connectivity is specified by selecting a Cytoscape network. In addition, each link must be assigned a weight and a direction through the query form. Edge weights are set using the \emph{Weight attribute} dropdown box, which lists all available floating-point edge attributes of the selected network and the default option (\emph{NONE}). If the default option is selected, \citm\ assumes a weight $2$ for any self-pointing edge and $1$ for all other edges. If an attribute is selected, the weight of an edge is set to the value of the selected attribute for that edge. Null attribute values are treated as zero weights.

\begin{figure}[ht!]
\begin{center}
\includegraphics{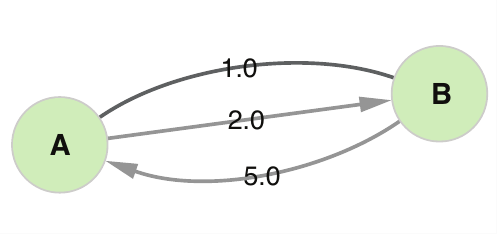}
\caption{\textbf{Edge weights example}
Consider the following example: Suppose A and B are nodes in a Cytoscape network linked by three edges of two types with shown edge weights. Assume two type I edges (lighter gray), $A\to B$ and $B\to A$ are directed, while a single type II edge (darker gray) $A\to B$ is undirected. At query time, \citm\ creates two directed edges, $A\to B$ and $B\to A$, with weights $3$ and $6$, respectively.
} \label{fig:edgeweights}
\end{center}
\end{figure}

Since Cytoscape edges are always internally treated as directed, the user must also indicate the directedness of each edge type through the query form. Whenever a new Cytoscape network is selected, \citm\ updates the query form and places all of the network's edge types into the \emph{undirected} category. The user can use arrow buttons to move some edge types to the \emph{directed} or \emph{ignored} category. Undirected edges are treated as bidirectional, with the same weight in both directions. Directed edges have a specified weight assigned only in the forward direction, with the backward direction receiving the zero weight. Ignored edges have zero weight in both directions. Since Cytoscape allows multiple edges of different types between the same nodes, \citm\ collapses multiple edges in each direction into a single edge by appropriately summing their weights (Fig.~\ref{fig:edgeweights}). 

\subsection*{Step two: selecting a model and boundary nodes}

In addition to a weighted directed graph, \itm\ requires an information flow model (emitting, absorbing or normalized channel), a selection of sources and/or sinks, and dissipation probability. The choice of the model determines the types of boundary nodes that need to be specified, as well as the ways in which the damping factor can be set (see `Step three: specifying dissipation probability' below). The query form also allows users to specify \emph{excluded nodes}. Any flow reaching excluded nodes is fully dissipated. This is a way to remove those nodes that do not participate in information propagation in the desired context or that introduce undesirable shortcuts.

\subsection*{Step three: specifying dissipation probability}

The values of $H$, $F$, and $\nPhiT{}{}{}$ all implicitly depend on the dissipation probability. In \itm\, the user can set the dissipation probability directly or specify a related quantity that can, using Newton's method, determine the dissipation probability. The choice of the alternative quantity depends on the selected model. For the emitting model, this quantity is the average path length before termination, which we denote by $\bar{t}$. For example, the user can require a random walker to make on average three steps before terminating. The formula for $\bar{t}$ is
\begin{equation}
\bar{t} = 1 + \frac{1}{n_S} \sum_s \sum_{j} H_{sj},
\end{equation}
where $n_S$ denotes the number of sources. For the normalized channel model, the path length before termination is given by 
\begin{equation}
\bar{t} = 1 + \frac{1}{n_S} \sum_s \sum_{j} \nPhiT{j}{s}{K}.
\end{equation}
Since the normalized channel model counts only the random walkers actually terminating at sinks, $\bar{t}$ is in this case bounded below by the length of the shortest path from any source to any sink. Hence, \itm\ accepts the desired value of $\bar{t}$ in terms of length deviation from the shortest path. There are two ways to set the average path-length deviation: in absolute units (steps) or as a proportion of the length of the shortest path. The absorbing model allows users to obtain the dissipation probability by setting the average absorption probability, denoted $\bar{r}$. The formula for $\bar{r}$ is
\begin{equation}
\bar{r} = \frac{1}{n_T} \sum_i \sum_k F_{ik},
\end{equation}
where $k$ ranges over all sinks, $i$ ranges over all transient nodes \emph{that are connected to at least one sink}, and $n_T$ is the total number of such nodes. The value of $\bar{r}$ represents the likelihood of a random walk starting at a randomly selected point in the network to reach a sink. The dissipation probability obtained in this way is larger if the sinks are well-connected hubs near the center of the network, in contrast to the case when the chosen sinks are not as well connected.

\subsection*{Step four: submitting a query}

After specifying all necessary input, the user submits a query by pressing the \emph{QUERY} button on the query form. The time required for a run depends on whether the query is local or remote, as well as on the size of the submitted graph and the number of selected sources and/or sinks.

%%%%%%%%%%%%%%%%%%%%%%%%%%%%%%%%
\section*{Output}

For every completed query, \citm\ displays its results in a viewer embedded in Cytoscape Results Panel and a new Cytoscape network consisting of significant nodes (ITM subnetwork). The results viewer has five tabs: \emph{Top Scoring Nodes}, \emph{Summary}, \emph{Input Parameters}, \emph{Excluded Nodes}, and \emph{Display Options}. The first four tabs contain information about the query and the results, while the last one contains a form that allows users to manipulate the ITM subnetwork. The form controls two aspects of the subnetwork: composition (what nodes are selected and how many) and node coloring.

\subsection*{Displaying significant nodes}

Subnetwork nodes are selected through a \emph{ranking attribute}, which assigns a numerical value from \itm\ results to each node. The nodes are listed in descending order of the ranking attribute and top nodes are displayed as the ITM subnetwork. The number of top nodes is determined by specifying a \emph{selection criterion}, which can be simply a number of nodes to show, a cutoff value or the `participation ratio'. Specifying a cutoff value $x$ selects the nodes with their ranking attribute greater than $x$. Participation ratio estimates the number of `significant' nodes by considering 
all values of the ranking attribute in a scale-independent  manner~\cite{SY07}. The available choices for the ranking attribute depend on the \itm\ model and the number of boundary points. For the emitting and normalized channel model, the user can select visits to a node from each source or the sum of visits from all sources. It is also possible to use \emph{interference}~\cite{SY07}, which denotes the minimum number of node visits, taken over all sources. For the absorbing model, the available attributes are absorbing probabilities to each sink and the total probability of termination at a sink. The values of all attributes for the subnetwork nodes are displayed in the \emph{Top Scoring Nodes} tab.

The colors of the subnetwork nodes are determined by selecting \emph{coloring attributes}, a \emph{scaling function} and a \emph{color map}. The list of coloring attributes is the same as the list of ranking attributes but the user can select up to three coloring attributes. If a single attribute is selected, node colors are determined by the selected eight-category ColorBrewer~\cite{HB03} color map. Otherwise, they are resolved by color mixing: each coloring attribute is assigned a single basic color (cyan, magenta or yellow), and the final node color is obtained by mixing the three basic colors in proportion to the values of their associated attributes at that node. The scaling function serves to scale and discretize the coloring attributes to the ranges appropriate for color maps. Figure~\ref{fig:models} shows examples of mixed color scheme with three boundary points (left and right columns) and of a coloring using a single attribute (center column).

\subsection*{Manipulating node attributes}

Since the \itm\ query results are saved as Cytoscape attributes of the original network, they can be arbitrarily modified through Cytoscape. Any changes made are reflected in the results viewer and the corresponding ITM subnetwork after pressing the \emph{RESET} button on the Display Options form. Using the \citm\ attribute nomenclature, users can create additional attributes to be used for ranking or coloring. Consider the following usage example. A user has run an emitting model query with three sources, \texttt{S1}, \texttt{S2}, and \texttt{S3}, and obtained the results in a viewer labeled \texttt{ITME243}. At the end of the run, \citm\ created the attributes \texttt{ITME243[S1]}, \texttt{ITME243[S2]} and \texttt{ITME243[S3]} for the nodes of the input network and saved the results as their values. The user creates a new floating-point node attribute with a label \texttt{ITME243[avgS1S2]} and fills it with an average of \texttt{ITME243[S1]} and \texttt{ITME243[S2]}. After resetting the Display Options form, an item `\texttt{Custom [avgS1S2]}' is available for selection as a ranking or coloring attribute. This gives the user the flexibility to reinterpret \texttt{S1} and \texttt{S2} as if they were a single source of equal weight as \texttt{S3}. Another possibility is to combine the results of queries with different boundaries and display them together on the same subnetwork.

\subsection*{Saving and restoring results}

The query network together with its attributes containing \itm\ results can be saved as a Cytoscape session and later retrieved. After reloading the session, the user can regenerate the results viewer and the corresponding subnetwork for a stored ITM by pressing the \emph{LOAD} button on the \citm\ query form and selecting the desired ITM from a list.

Alternatively, the \itm\ results can be exported to tab-delimited text files through the Cytoscape \emph{Export} menu. Each exported tab-delimited file contains all the information necessary to restore the results except the query network and can be easily manipulated both by humans and by external programs or scripts. The results from tab-delimited files can be imported into any selected Cytoscape network through the \emph{Import} menu. Since the selected network may be different from the original query network, only the results for the nodes in the selected network whose IDs match the IDs from the imported file will be loaded. After importing the results, \citm\ generates a new results viewer and a subnetwork, as if the results originated from a direct \itm\ query.

\section*{Discussion}

The main function of \itm, also applicable to domains other than PPI networks, is to retrieve information from large and complex networks by discovering the possible interface between network nodes that are hypothesized to be related. This paradigm works best with large networks, where such information cannot be easily accessed by other means. For examples of applications of the \itm\ frameworks to protein-protein interaction networks, consult our earlier papers~\cite{SY07,SY09b,SY10b}.

With a network as an \emph{encyclopedia} of domain-specific knowledge, \itm\ enables a direct access to its specific portions related to a specified context. The user can learn about the objects representing individual nodes by setting them as sources and/or sinks and retrieving information about the most significant objects in the resulting ITM. This approach not only extracts a relevant subnetwork but also produces context-specific weights for each node. With their interpretation as average numbers of node visits, or equivalently, as average numbers of paths passing through a node, the ITM weights signify the relative importance of network nodes in the context of the query and thus can be used to refine its interpretation as a whole. For example, a single node with a large weight in an ITM resulting from a normalized channel model query represents a choke point \emph{in the particular context of the query}. The same node need not have a high global centrality. 

Containing both sources and sinks, the normalized channel model offers the users the ability to formulate and evaluate network based hypotheses \textit{in silico}. Since information flow that reaches one sink cannot subsequently terminate at any other, sink nodes can be associated with alternative hypotheses, such as different biological functions if the network is PPI. The information flow from each source will then, depending on the dissipation coefficient used, mainly trace the path towards the sink most likely to be reached first from that source (see Fig.~\ref{fig:models}, right column). The \itm\ framework considers all weighted paths from sources to sinks and hence produces more robust results than approaches involving only the shortest paths. The path weights are tunable using the dissipation probability.

Compared to the previously described web interface to \itm~\cite{SY09b}, \citm\ significantly benefits from being a part of the Cytoscape platform. Although the \emph{Display Options} form is very similar to the web version, the sophisticated network visualization functionality provided by Cytoscape allows significantly more versatility in displaying ITMs. % than that is possible by using Graphviz programs. 
For example, Cytoscape GUI allows users to manually alter node placements, rotate network views, or arbitrarily change the look of a network. In addition, Cytoscape interface enables users to directly manipulate node attributes representing \itm\ results and possibly create new node summary variables appropriate to their problem. The newly created variables can be immediately reflected in the graphical representation of an ITM, which is not possible in the web setting. Most importantly, the results of \itm\ can be integrated into workflows involving other Cytoscape plugins that provide complementary functionality. For instance, output ITMs can be related to terms from controlled vocabularies such as Gene Ontology~\cite{ABBB00} using functional enrichment analysis plugins such as PinGO~\cite{SOIM11} or our own recently released \cssm~\cite{SBY12}. The graph-theoretic structure of ITM subnetworks can be analyzed using a variety of algorithms such as MCODE~\cite{BH03} or GraphletCounter~\cite{WS2012,MP08}. 

The architecture of \citm\, with a Cytoscape front end and an \itm\ back end offers flexibility for a variety of usage scenarios. In contrast to the web version, it allows users to use \itm\ with arbitrary networks and edge weights, rather than being limited to compiled PPIs from few model organisms. Most users will be content with accessing \itm\ through the web server. However, the option to download and install the \qmbpmnt\ package  provides not only faster running times for queries but also the ability to use the command line interface for \itm\, to perform batch queries and to locally reproduce its web service. The separation of the presentation layers (web or Cytoscape) from the `business' layer (standalone \itm) facilitates easy future updates to any components.

%%%%%%%%%%%%%%%%%%%%%%%%%%%%%%%%
\section*{Conclusion}

\citm\ is a plugin that brings the previously unavailable network flow algorithms of \itm\ to the Cytoscape platform. It enables users to extract context-specific subnetworks from large networks by specifying the origins and/or destinations of information flow. \citm\ significantly extends the features of the previously released web version of \itm. 

The main novelty of \citm\ is that it allows the user to specify as input any Cytoscape network, rather than being restricted to the PPI networks available through the \itm\ web service. Using Cytoscape attributes to hold their desired values, users may easily supply their own edge weights and denote edge directionalities. Additionally, the ability to manipulate and add new node attributes through Cytoscape reduces the workload required for visualizing various combinations of ITM components. In the context of biological cellular networks, this additional flexibility may lead to constructions of new node attributes that can better reflect biological significance, hence facilitating more educated hypothesis forming. 

By bringing \itm\ to Cytoscape, \citm\ enables seamless integration of \itm\ results with other Cytoscape plugins having complementary functionality for data analysis. By decoupling the query network from the information flow algorithm, the newly developed \citm\ can be applied to many other domains of network-based research beyond protein-networks.

%%%%%%%%%%%%%%%%%%%%%%%%%%%%%%%%
\section*{Availability and requirements}

\subsection*{\citm\ plugin}
\textbf{Project name:} CytoITMprobe \\
\textbf{Project home page:} \url{http://www.ncbi.nlm.nih.gov/CBBresearch/Yu/downloads/itmprobe.html} \\
\textbf{Documentation:} \url{http://www.ncbi.nlm.nih.gov/CBBresearch/Yu/mn/itm_probe/doc/cytoitmprobe.html} \\
\textbf{Video tutorial:} \url{http://www.youtube.com/watch?v=4Cdf-mSKtWo} \\
\textbf{Operating system(s):} Platform independent\\
\textbf{Programming language:} Java \\
\textbf{Other requirements:} Java SE 6 or higher and Cytoscape 2.7 or higher \\
\textbf{License:} All components written by the authors at the NCBI are released into Public Domain. Components included from elsewhere are available under their own open source licenses and attributed in the source code.

\subsection*{Standalone \itm\ (optional for \citm)}
\textbf{Project name:} qmbpmn-tools \\
\textbf{Project home page:} \url{http://www.ncbi.nlm.nih.gov/CBBresearch/Yu/downloads/itmprobe.html} \\
\textbf{Documentation:} \url{http://www.ncbi.nlm.nih.gov/CBBresearch/Yu/mn/itm_probe/doc/} \\
\textbf{Operating system(s):} Platform independent\\
\textbf{Programming language:} Python \\
\textbf{Other requirements:}  Python 2.6 or 2.7, Numpy 1.3 or higher and Scipy 0.7 or higher. UMFPACK Scikit is recommended for good performance.\\
\textbf{License:} All components written by the authors at the NCBI are released into Public Domain. Components included from elsewhere are available under their own open source licenses and attributed in the source code.

\section*{Acknowledgments}
This work was supported by the Intramural Research Program of the National Library of Medicine at the National Institutes of Health.

%\bibliography{cytoitmprobe}
%\bibliographystyle{abbrvnat}

\end{document}